\documentclass[%
 aip,
 rsi,%
 amsmath,amssymb,
preprint,%
]{revtex4-1}

\usepackage{graphicx}
\usepackage{dcolumn}
\usepackage{bm}
\usepackage{natbib}

\begin{document}{}{}

\title{Local imaging of high mobility two-dimensional electron systems with virtual scanning tunneling microscopy}

\author{M. Pelliccione} 
\affiliation{Department of Applied Physics, Stanford University, 348 Via Pueblo Mall, Stanford, CA 94305}
\affiliation{Stanford Institute for Materials and Energy Sciences, SLAC National Accelerator Laboratory, 2575 Sand Hill Road, Menlo Park, CA 94025}
\affiliation{Department of Physics, University of California, Santa Barbara, Santa Barbara, CA 93106}
\author{J. Bartel}
\affiliation{Stanford Institute for Materials and Energy Sciences, SLAC National Accelerator Laboratory, 2575 Sand Hill Road, Menlo Park, CA 94025}
\affiliation{Department of Physics, Stanford University, 382 Via Pueblo Mall,  Stanford, CA 94305}
\author{A. Sciambi}
\affiliation{Department of Applied Physics, Stanford University, 348 Via Pueblo Mall, Stanford, CA 94305}
\affiliation{Stanford Institute for Materials and Energy Sciences, SLAC National Accelerator Laboratory, 2575 Sand Hill Road, Menlo Park, CA 94025}
\author{L. N. Pfeiffer}
\affiliation{Department of Electrical Engineering, Princeton University, Princeton, New Jersey 08544}
\author{K. W. West}
\affiliation{Department of Electrical Engineering, Princeton University, Princeton, New Jersey 08544}
\author{D. Goldhaber-Gordon}
\affiliation{Stanford Institute for Materials and Energy Sciences, SLAC National Accelerator Laboratory, 2575 Sand Hill Road, Menlo Park, CA 94025}
\affiliation{Department of Physics, Stanford University, 382 Via Pueblo Mall,  Stanford, CA 94305}

\date{\today}

\begin{abstract}
Correlated electron states in high mobility two-dimensional electron systems (2DESs), including charge density waves and microemulsion phases intermediate between a Fermi liquid and Wigner crystal, are predicted to exhibit complex local charge order. Existing experimental studies, however, have mainly probed these systems at micron to millimeter scales rather than directly mapping spatial organization. Scanning probes should be well-suited to study the spatial structure of these states, but high mobility 2DESs are found at buried semiconductor interfaces, beyond the reach of conventional scanning tunneling microscopy. Scanning techniques based on electrostatic coupling to the 2DES deliver important insights, but generally with resolution limited by the depth of the 2DES. In this Letter, we present our progress in developing a technique called ``virtual scanning tunneling microscopy'' that allows local tunneling into a high mobility 2DES. Using a specially-designed bilayer GaAs/AlGaAs heterostructure where the tunnel coupling between two separate 2DESs is tunable via electrostatic gating,
combined with a scanning gate, we show that the local tunneling can be controlled with sub-200 nm resolution. \end{abstract}

\maketitle

\section{\label{sec:introduction}INTRODUCTION}
A local tunnel probe can complement traditional transport measurements for understanding mesoscopic systems~\citep{Eriksson1996,Martin2007,Tessmer1998,Topinka2001,Jura2007,Hackens2010,Huber2008,Gildemeister2007,Pelekhov1998,Moussy2001,Crook2002,Hedberg2010,Vinante2011,Song2010,Pan1999a,Brown2004,Zhang2011,Urazhdin2000,Pelekhov1999}. In many situations where electron-electron interactions are important, non-trivial spatial ordering of the electrons is a predicted consequence that cannot be fully probed through transport \citep{Pothier1997}. For example, quantum Hall stripes are predicted to form at half-integer filling factors due to a characteristic anisotropy in transport along different crystalline axes in GaAs \citep{Lilly1999}. Anisotropic transport has been measured in this regime, \citep{Fradkin1999} likely because of the charge density wave (CDW) associated with these stripes. Scanning tunneling microscopy (STM) has been used extensively to study CDW behavior \citep{Wise2008,Brun2010,Coleman1988}, especially in the context of high-temperature superconductivity, but the buried nature of high mobility 2DESs precludes this type of measurement.

\begin{figure}
\centering
\includegraphics[scale=1, viewport=0 0 350 350]{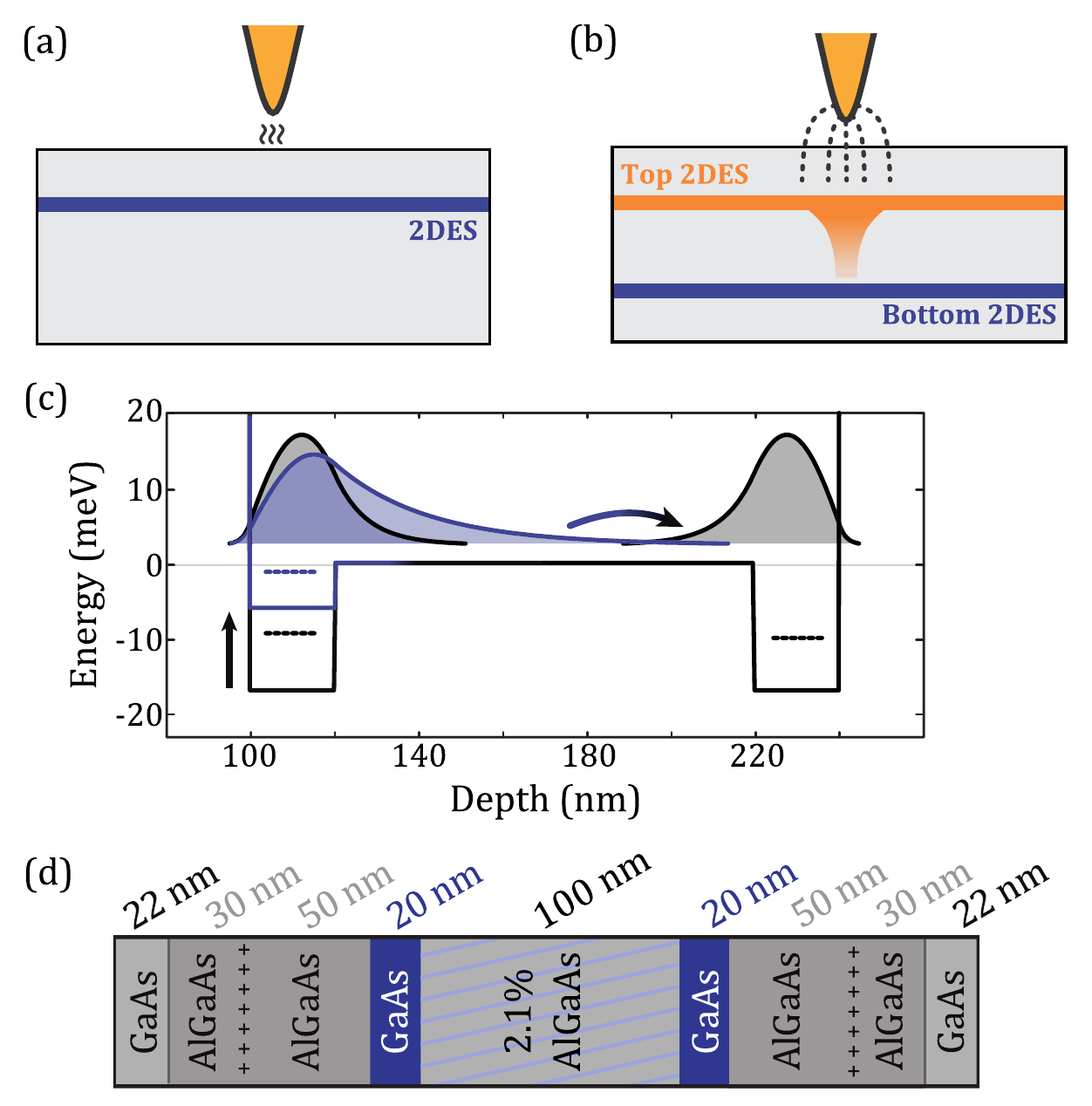}
\caption{Comparison between (a) performing STM on a buried 2DES, and (b) performing VSTM on a bilayer 2DES. In (a), the surface energy barrier coupled with the depth of the 2DES results in a small tunnel current from an STM tip. In (b), the presence of the tip locally gates the heterostructure and induces tunneling between the top and bottom 2DESs preferentially beneath the tip. (c) Diagram of the wavefunction extension effect, which is used to modulate tunnel couplings with an applied electric field. Reducing the electron density in one quantum well results in a larger wavefunction overlap between states localized in the two quantum wells, leading to increased tunneling. (d) VSTM sample heterostructure. Two 20 nm wide quantum wells are located 100 nm from each surface, with a 2.1\% aluminum barrier between them.}
\label{fig:stmvstm}
\end{figure}

The work presented in this Letter is aimed at developing a technique that can overcome the obstacles to using traditional STM on mesoscopic devices, as depicted in Figure \ref{fig:stmvstm}. The key idea is to change the nature of the tunnel barrier, from the surface energy barrier which is fixed by material properties, to a tunable tunnel barrier inside a semiconductor heterostructure that can be modulated by an electric field. Taking advantage of the flexibility of the GaAs/AlGaAs materials system, one can grow heterostructures that support two separate 2DESs \citep{Sciambi2010}, and by controlling the aluminum doping define the height of the tunnel barrier between them. The tunneling is controlled locally by scanning a sharp metal tip above the heterostructure, which serves to electrostatically change the height of the tunnel barrier and enhance tunneling preferentially beneath the tip. Therefore, instead of tunneling from a 3D metal into a 2DES as would be the case in STM, the tunneling takes place between two interfacial 2D systems. This technique is called ``virtual scanning tunneling microscopy'' (VSTM) because a ``virtual'' STM tip is induced inside the heterostructure between the two 2DESs.

A schematic depicting the tunneling process in a VSTM device is shown in Figure \ref{fig:stmvstm}(c). As the electron density is lowered in one of the 2DESs with an electric field, the subband energy will become less negative relative to the Fermi energy, and therefore the difference between the subband energy and the conduction band edge in the barrier region will be reduced. This implies that as the density is reduced, the tunneling conductance should \textit{increase}. This effect is called wavefunction extension because the tail of the wavefunction for electrons in one quantum well extends further into the barrier as the density is reduced, thereby leading to a stronger overlap with the wavefunction localized in the other quantum well, and increasing tunneling.

The layer design for the heterostructure investigated in this work is shown in Figure \ref{fig:stmvstm}(d). Two 2DESs are formed in two 20 nm wide GaAs quantum wells approximately 100 nm from the top and bottom surfaces of the device. The barrier region
has 2.1\% aluminum doping, bringing the conduction band approximately 1 meV above the Fermi energy in this region. The 2DESs have densities of $n = (2,3) \times 10^{11}$ cm$^{-2}$ with mobilities of $\mu = (1,3) \times 10^6$ cm$^{2}$V$^{-1}$s$^{-1}$ at 3.3 K, respectively. Large-area tunneling and lateral transport in this heterostructure was reported previously \citep{Sciambi2010,Sciambi2011}. The tunneling process was shown to scatter in-plane momentum in these devices, which allows for tunneling between 2DESs with different electron densities. To perform scanning measurements, the active GaAs/AlGaAs layer is thinned to approximately 400 nm using an epoxy bond and stop etch (EBASE) procedure \citep{Weckwerth1996}, which allows Ti/Au gates to be placed in close proximity to both quantum wells.

\section{DEMONSTRATION OF VSTM}
\begin{figure}
\includegraphics[scale=0.75, viewport=0 0 500 225]{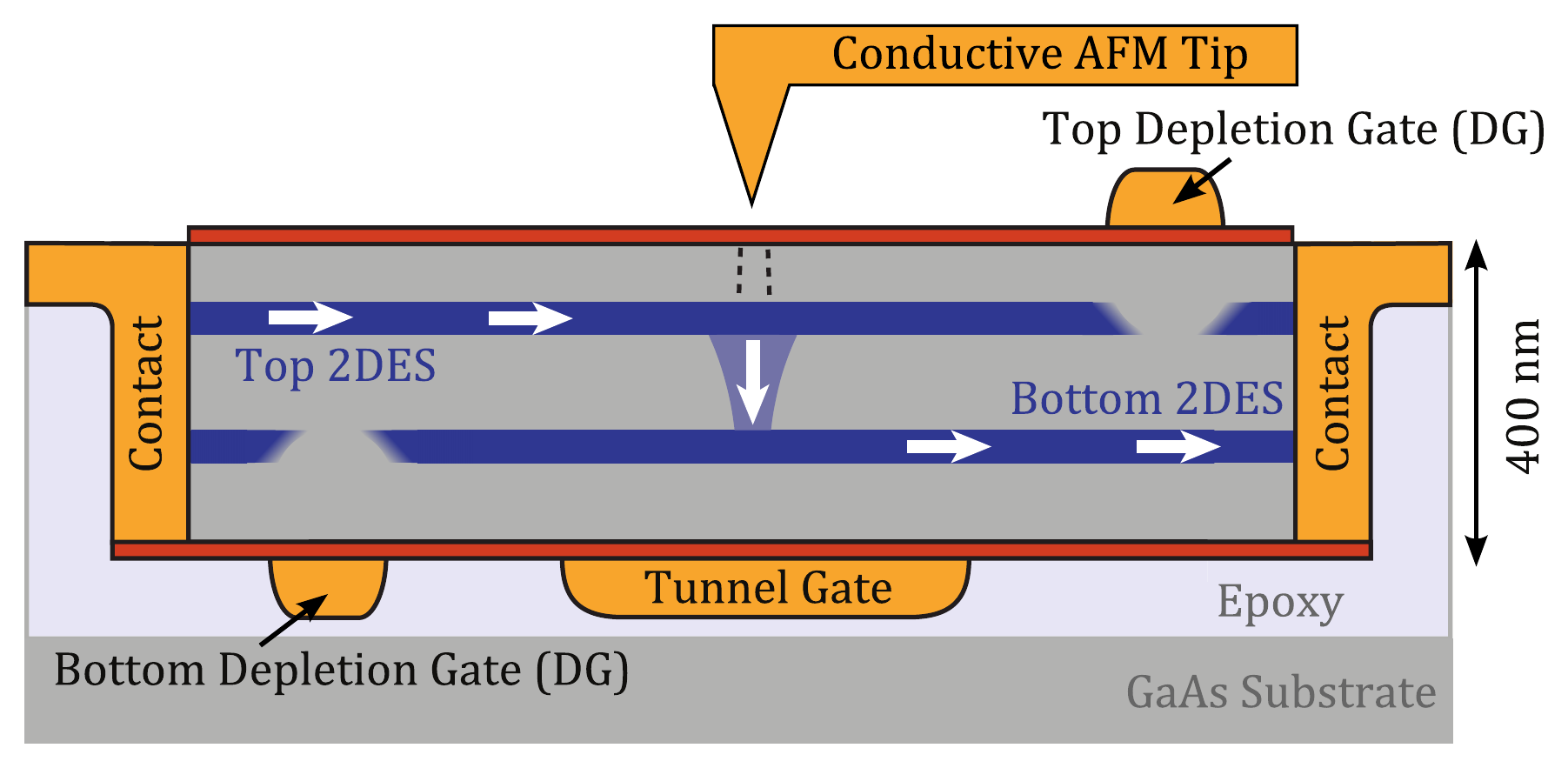}
\centering
\caption{Device schematic for VSTM measurements. A metallized AFM tip locally gates the device to enhance the tunnel conductance between the top and bottom 2DES. The white arrows depict the direction of current flow though the device.}
\label{fig:vstmschem}
\end{figure}

\begin{figure}
\includegraphics[scale=0.5, viewport=0 0 575 1100]{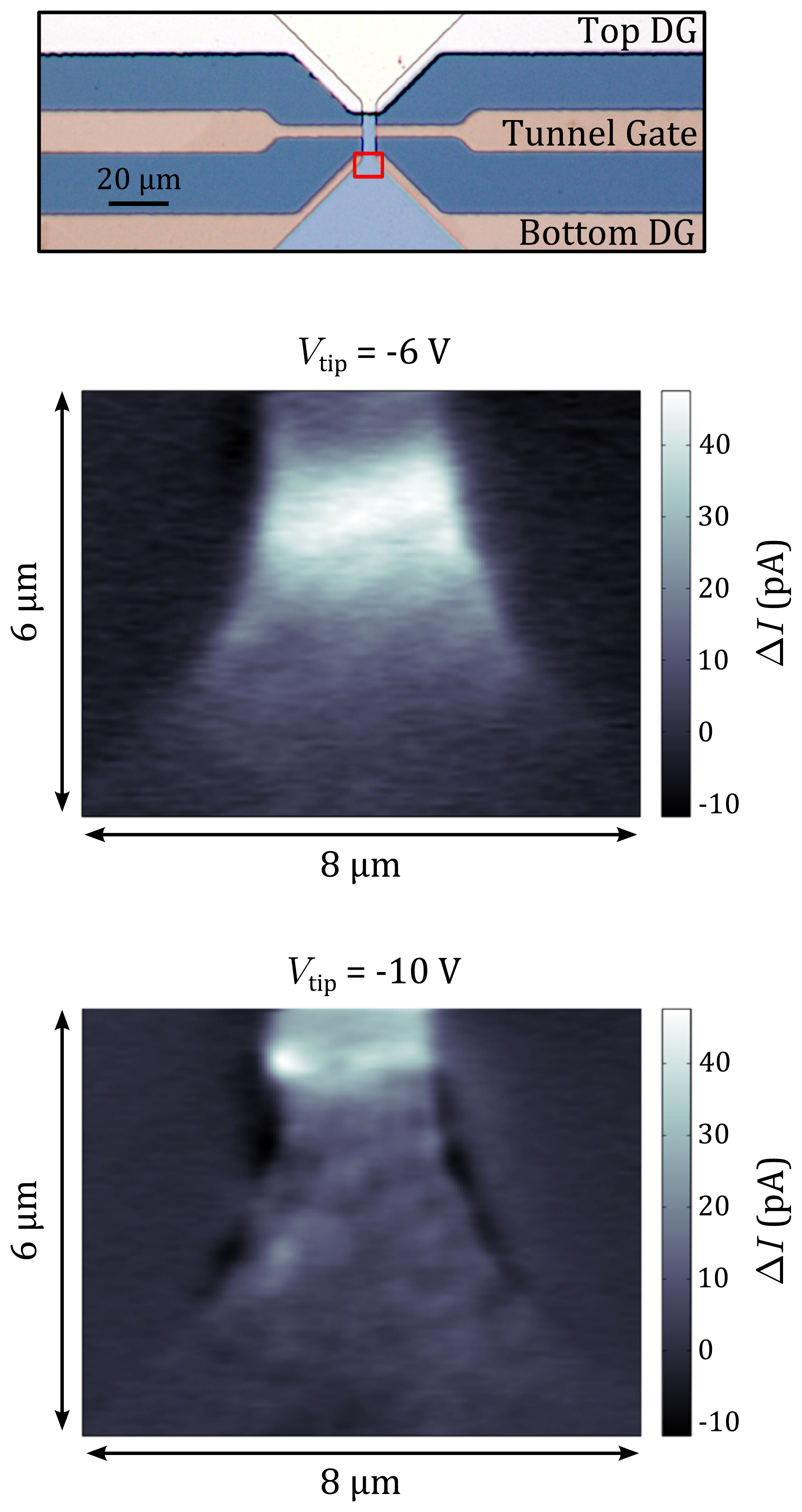}
\centering
\caption{Realization of VSTM. Scans over the red bounded region in the optical image measure induced tunneling versus tip position for two tip voltages, $-6$ V and $-10$ V. Bright areas in the images indicate strong tunneling induced locally by the AFM tip. The tip height is 300 nm during the measurement, with a 30 nm RMS height oscillation.}
\label{fig:vstmimg}
\end{figure}

A schematic of the device geometry used for VSTM measurements is shown in Figure \ref{fig:vstmschem}. 
Depletion gates (DG) located on the top and bottom of the device are biased such that one ohmic contact connects to the bottom 2DES and one to the top 2DES~\citep{Eisenstein1990a}. Therefore, when conductance is measured between the ohmic contacts, current must flow by tunneling in the region between the depletion gates. The ``tunnel region'' bounded by the depletion gates is gated from the bottom by the tunnel gate. There is no large gate on the top of the device over the tunnel region, instead this region is gated by a scannable metallized AFM tip. The active mesa region between the top and bottom depletion gates is 5 $\mu$m wide and 15 $\mu$m long, of which the middle 5 $\mu$m is gated by the tunnel gate, as shown in the optical micrograph in Figure \ref{fig:vstmimg}. Not pictured are marks deposited elsewhere on the chip with a 20 $\mu$m grid spacing to help locate the device via AFM topography. The measurements are carried out in a home-built scanning gate microscope housed in a cryogen-free dilution refrigerator \citep{Pelliccione2013}. Although the system is capable of achieving electron temperatures as low as 45 mK, for the discussion that follows all measurements are taken at 3.3 K.

Returning to the schematic of the measurement shown in Figure \ref{fig:vstmschem}, we would like to determine the tunnel current induced by the tip, but there is also a tunnel current from the areas of the device not gated by the tip. Though this current is smaller per unit area, the area bounded by the depletion gates is 5 $\mu$m $\times$ 15 $\mu$m, much larger than the area gated by the tip, so this background tunneling is significant. To overcome this, we oscillate the height of the tip and measure the tunnel conductance response at the oscillation frequency. This serves to reduce the background large-area gating due to metal on the cantilever. Oscillating the height of the tip rather than the voltage on the tip leads to crisper imaging results. If we oscillate the tip voltage, metal on the cantilever capacitively modulates tunneling over a large area, whereas if we oscillate the tip height the tunneling is strongly modulated only where the tip is closest to the device.

We can now start analyzing the response of the device to local gating. In Figure \ref{fig:vstmimg}, the induced tunneling over the red bounded region in the optical micrograph is measured under two different gating conditions, $-6$ V on the tip and $-10$ V on the tip. In both cases, the tip height is 300 nm above the surface with a 30 nm RMS height oscillation, the tunnel gate is set at $-490$ mV, and an interlayer bias of 15 mV is applied to the device. For this device geometry, a signal is measured only when the tip has induced tunneling locally in the device. The response of the device changes significantly under the different gating conditions. In both scans, there is a region of strongest induced tunneling, where $\Delta I > 40 $ pA. This region represents the gating condition where the device is being operated at maximum gain. This region shifts upward in the image from $-6$ V to $-10$ V due to the effect of the bottom depletion gate, which is underneath the device over the bottom half of the images.

The data for $V_{\rm tip} = -10$ V demonstrate the device response when it is ``over-gated'', meaning the top 2DES has been depleted past the mobility edge. Comparing the width of the mesa as seen in both images, it is narrower in the $-10$ V image and there is even a region near the edges where the response is negative, implying that the top 2DES is being totally depleted at the edges. In addition, an emergent fine structure is visible in the lower half of the image, a consequence of the non-uniform potential in the top 2DES in the presence of disorder.

\section{SPATIAL RESOLUTION}

\begin{figure}
\includegraphics[scale=0.8, viewport=0 0 550 200]{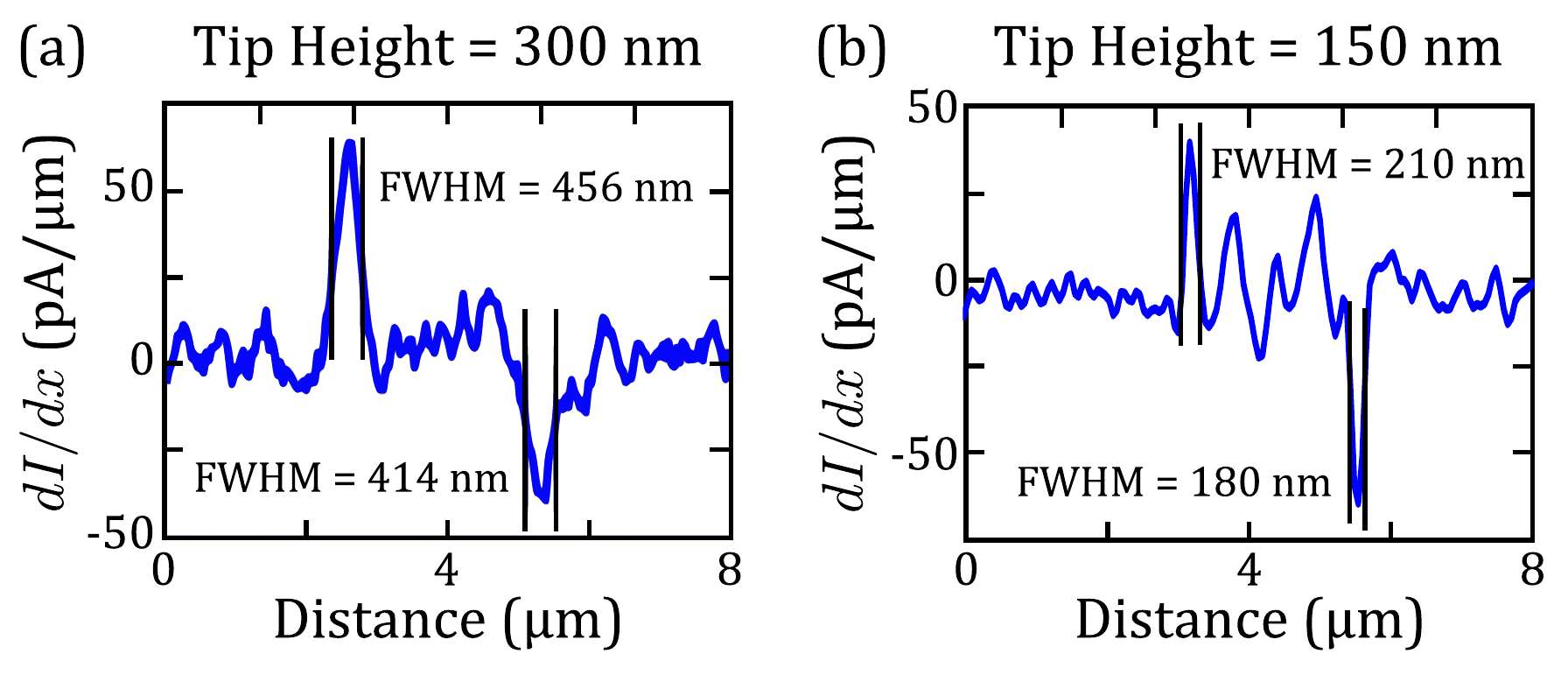}
\centering
\caption[Spatial resolution of VSTM.]{The lateral resolution of the measurement can be estimated by measuring how sharp the edge of the device appears in the data at a tip height of (a) 300 nm ($V_{\rm tip} = -6$ V) and (b) 150 nm ($V_{\rm tip} = -4$ V) above the surface. The full width at half maximum (FWHM) of the derivative along a line cut gives a measure of the sharpness of the mesa, which improves from 435 nm to 195 nm as the tip is moved closer to the device.}
\label{fig:vstmres}
\end{figure}

One important parameter that can be extracted from the data in Figure \ref{fig:vstmimg} is the lateral resolution of the measurement. We can estimate the lateral resolution by measuring how sharp the edges of the mesa appear in the tunneling data. This is accomplished by taking a line cut perpendicular to the mesa and taking a derivative to measure the full width at half maximum (FWHM) of the peak at the edge of the mesa. This analysis is shown in Figure \ref{fig:vstmres} for two different tip heights, 300 nm and 150 nm above the surface.

As expected, the resolution improves as the tip is brought closer to the device since the electric field from the tip becomes sharper as the tip is moved closer. The average FWHM improves from 435 nm to 195 nm as the tip is brought closer, but the trace in Figure \ref{fig:vstmres}(b) indicates that there are new features in the data (around a distance of 4 $\mu$m) not seen in Figure \ref{fig:vstmres}(a). These features are due to a combination of surface contamination and charge traps in the dielectric on the surface, challenges which must be addressed in the device fabrication before pushing the measurements to a higher resolution. A higher resolution should be achievable, since electrostatic simulations published previously \citep{Sciambi2010} predict a spatial resolution of 40 nm. The spatial resolution required for the technique depends on the specific system to be studied, but one can use the Fermi wavelength at a given electron density as a suitable metric. At an electron density of $5 \times 10^{10} \mbox{ cm}^{-2}$, which can be achieved in current devices before hitting the mobility edge, the Fermi wavelength is $\lambda_F = \sqrt{2\pi/n} = 112$ nm. To observe Friedel oscillations associated with electron scattering off defects and edges in the 2DES, as have been studied with STM for surface 2DESs \citep{Crommie1993,Binnig1983,VanderWielen1996,Hasegawa1993,Sprunger1997}, one would require a spatial resolution smaller half the Fermi wavelength, or 56 nm in this example.

\section{CONCLUSION}
The discussion presented in this Letter demonstrates the viability of VSTM as a powerful scanning probe technique to study mesoscopic electron systems. We have shown that the tunneling between two buried 2DESs can be modulated locally with a scanning gate on a length scale below 200 nm, but need remains to optimize the device design and fabrication procedure to achieve the expected spatial resolution of 40 nm. The future directions for VSTM are very broad, and in the near term can include measurements of local tunneling into quantum Hall edge states, along with observing the spatially oscillatory electron density of states near impurities and edges. Also, by varying the interlayer bias applied to a VSTM device, the energy dependent density of states can be measured locally, in analogy to STM. More generally, the VSTM technique can be extended to any interfacial materials system where the tunnel barrier can be engineered through growth. 

\begin{acknowledgments}
The authors acknowledge support by the Department of Energy, Office of Basic Energy Sciences, under contract DE-AC02-76SF00515, with the original concept developed under the Center for Probing the Nanoscale (NSF NSEC Grant No. 0425897) and a Mel Schwartz Fellowship from the Stanford Physics Department. M.P. acknowledges support from the Hertz Foundation, NSF, Stanford and Intel.
\end{acknowledgments}
%

\end{document}